\documentclass[showpacs,twocolumn,superscriptaddress,nofootinbib]{revtex4}

\usepackage{graphicx}
\usepackage{dcolumn}
\usepackage{bm}
\usepackage{color}

\usepackage{amsmath}
\usepackage{amssymb}


\begin{document}

\title{Maximum Force for Black Holes and Buchdahl Stars }

\author{Naresh Dadhich}
\email{nkd@iucaa.in}

\affiliation{Inter University Centre for Astronomy \&
Astrophysics, Post Bag 4, Pune 411007, India }

%\date{\today}
\begin{abstract}
Black hole and Buchdahl star are identified respectively by $\Phi(R)=1/2, 4/9$ where $g_{tt}=1-2\Phi(R)$ for a spherically symmetric static metric. We investigate the maximum force for black hole and Buchdahl star when one of the participating objects  is charged and/or rotating while the other is neutral and non-rotating. It turns out that the maximum force between two Schwarzschild objects is universal, given in terms of the fundamental constant velocity of light and the gravitational constant in general relativity (GR) in the usual four dimensional spacetime. In general this feature uniquely picks out the pure Lovelock gravity (having only one $N$th order term in action which includes GR in the linear order $N=1$) and the dimensional spectrum, $D=3N+1$, where $N$ is degree of the Lovelock polynomial action.
\end{abstract}
 \maketitle

\section{Introduction}
\label{introduction}

In Newtonian gravity, there is no upper bound on force which goes on increasing as distance between the objects decreases, and in the limit of point particles it diverges. In GR on the other hand as body becomes more and more compact it turns into a black hole --- the black hole horizon forms that provides the lower bound on the separating distance. It could then define the maximum value a force between any two objects can physically attain. It has been computed in various situations, the maximum force or tension between two equal mass static uncharged Schwarzschild black holes touching each-other at the horizon \cite{Gibb2002, JBGib} is given by
\begin{equation}
F_{\max }= c^4/4G
\end{equation}
where $c$ is the velocity of light and $G$ is the Newtonian gravitational constant. This would in turn define the maximum power attainable in any physical system as
\begin{equation}
P_{\max }=cF_{\max }=c^5/4G\, .
\end{equation}
This is known as the so-called Dyson Luminosity \cite{dyson}, \textit{\ }or some multiple of it to account for geometrical factors $O(1)$. This limits maximum possible luminosity in gravitational or indeed any other forms of
radiation that an isolated system may emit, \cite{schiller}, \cite{sperhake}. Schiller has further proposed and surmised that the existence of a maximum force implies GR\footnote{This is though not exactly true as it turns out that pure Lovelock theory as well as Moffat's gravity theory \cite{moffatt} do admit maximum force. What is however true is the fact \cite{JB-ND} that existence of maximum force bound in terms of only the fundamental constant velocity of light and the gravitational constant, does lead to pure Lovelock gravity in $D=3N+1$, which includes GR at the linear order $N=1$ where $N$ is degree of the Lovelock polynomial action \cite{Love}.} just as maximum velocity characterises special relativity. Recently Schiller \cite{schiller21} has also proposed tests for the maximum force and power. There have been several computations of maximum force in different settings \cite{ong, ata} and also of maximum entropy emission \cite{mirza}. All this would have important implications on the cosmic censorship conjecture \cite{Barrow2004, Barrow2004a, Barrow2004b}.\\

Further let's write the maximum force in the Planck unit in $D$ spacetime dimensions which would read as
\begin{equation}
F_{pl}=G_{D}^{2/(2-D)}c^{(4+D)/(D-2)}h^{(4-D)/(2-D)},  \label{force}
\end{equation}
which is free of the Planck's constant $h$ if and only if $D=4$. In four spacetime dimension, it is solely given in terms of the fundamental constant velocity of light and the gravitational constant, and is free of everything else. \\

 Thus the bound is universal except for the dimensionality of spacetime and hence it is expected to remain true even when quantum gravity effects are included or in full theory of quantum gravity. It also turns out that it remains unaltered when the cosmological constant $\Lambda$ is included \cite{JBGib}.\\

Further also note that the ratio of magnetic moment to angular momentum is also free of the Planck's constant $h$ \cite{BarGib2}. This indicates something fundamental and natural about these non-quantum universal units that is unique to four dimensional
spacetime \cite{JBsing}. This bound however does not exist in the
Newtonian gravity \cite{JBnewt}, where point masses can approach arbitrarily close to each-other and the inverse-square gravitational force can then become arbitrarily large. It is the formation of an event horizon around these mass points in GR that is responsible for the the maximum force bound. It is the inverse square law that makes it universal, free of black hole masses, and only given in terms of $c$ and $G$. This is how the dimensionality of spacetime, $D=4$ gets singled out in GR, and in general $D=3N+1$ in pure Lovelock theory which includes GR for $N=1$ \cite{sum-dad}.\\

In astrophysics and cosmology besides black holes which mark the limiting configuration, there are other compact objects of interest and consequence like massive compact stars or large scale configurations. For a static object with perfect fluid interior there is well-known Buchdahl compactness bound \cite{buch} which is obtained for a fluid distribution under very general conditions that density and isotropic pressure are positive and the former is decreasing outwards, and it is matched at the boundary with the Schwarzschild vacuum metric. The bound turns out to be
\begin{equation}
M/R \leq 4/9 \, .
\end{equation} \\

Let's identify with the limiting value, equality, the Buchdahl star; i.e. $M/R = 4/9$. Note that for black hole, the compactness limit is  $M/R = 1/2$ which is absolute and unique. This is because its boundary is null, defining event horizon and hence is completely immune to other conditions. Black hole with a null boundary is naturally the most compact object. On the other hand Buchdahl star is the most compact non-horizon object with timelike boundary. Buchdahl bound can therefore be neither absolute nor unique. It can vary with the choice of equation of state, anisotropy of pressure and energy conditions \cite{andrea08, kara-stal08, mak-stuch, andrea09}. In a very recent interesting and insightful paper \cite{alho}, it has been shown that for realistic physical  conditions of causality and stability of fluid interior, Buchdahl bound is indeed always respected.  \\

Since Buchdahl star is the non black hole most compact object, it should be pertinent and interesting to investigate maximum force between two such objects. That is what we wish to study in this paper. In addition we would also investigate maximum force between two, one of which is charged and/or rotating black hole or Buchdahl star while the other is neutral and non-rotating. Further this investigation of maximum force for black holes and Buchdahl stars would be carried over to pure Lovelock gravity. In particular it turns out that the maximum force for Buchdahl star is $8/9$-th of the maximum force for black hole. \\

It is remarkable that existence of maximum force between two black holes or Buchdahl stars, and given entirely in terms of the fundamental constant velocity of light and the gravitational constant, uniquely singles out the pure Lovelock theory in dimensional spectrum $D=3N+1$ which includes GR for $N=1$ \cite{sum-dad}. We have considered force  between two equal mass Schwarzschild black holes with their horizons touching. If the masses are not equal then force is bounded by the maximum bound obtained for equal masses \footnote{If masses are unequal, $F = (M^{\prime}/M)F_{max} < F_{max}$ where $F_{max} = c^4/4G$.} \cite{JBsing}.

%In this paper we would study the maximum force bound for black holes with charge and rotation, and also for their Buchdahl counterparts; i.e., Buchdahl stars with charge and rotation. It is remarkable that the expressions for the maximum force in the two cases turn out to be parallel and similar, $F_{max}(Buch) = (8/9)^2 F_{max}(BH)(\alpha^2 \to (8/9)\alpha^2, \beta^2 \to (8/9)^2\beta^2)$ This is because $(M/R)_{Buch} = (8/9)(M/R)_{BH} (\alpha^2 \to (8/9)\alpha^2, \beta^2 \to (8/9)^2 \beta^2)$.\\

The paper is organized as follows: In the next section we shall characterize black holes and Buchdahl stars in terms of potential $\Phi(R)$ for radial motion, which in the static case is $g_{tt} = 1 - 2\Phi$. Then black hole and Buchdahl star are always defined by $\Phi = 1/2, 4/9$ respectively. In Sec. III we study the maximum force for charged and/or rotating objects, black hole or Buchdahl star. Sec IV is concerned with the pure Lovelock theory and investigation of maximum force for pure Lovelock objects. We conclude with Discussion. \\

\section{Black holes and Buchdahl stars}

A static object is described by spherically symmetric metric,
\begin{equation}
ds^2 = f(R) dt^2 - dR^2/f(R) - R^2d\Omega^2
\end{equation}
where $d\Omega^2$ is the metric on the unit sphere. Let's write $f(R)=1-2\Phi(R)$, the above metric describes neutral Shwarzschild object when $\Phi(R)=M/R$ and a charged Reissner-Nordstr{\"o}m one for $\Phi(R)=(M-Q^2/2R)/R$. \\

Black holes and Buchdahl star are respectively identified by the condition $\Phi(R) = 1/2, 4/9$. There is also another insightful characterization \cite{dad19}, black hole horizon is defined when gravitational field energy is equal to non-gravitational energy, and when it is half of non-gravitational energy defines Buchdahl star.  \\

A black hole is defined by its event horizon which would be given by $\Phi(R)= 1/2$. For a static Schwarzschild object, we know, $\Phi(R) = M/R$ and $\Phi(R) = 1/2$ gives the black hole horizon, $R=2M$. On the other hand Buchdahl star is identified by $\Phi(R) = M/R = 4/9$. So black hole and Buchdahl star are characterized by  $\Phi(R) = 1/2, 4/9$ respectively, and this is universal and holds good in general for charged and rotating objects as well. Further also note that $(M/R)_{Buch}/(M/R)_{BH}  = 8/9$ is also universal \cite{sum-dad22}. We shall generally work in the units with $G=c=1$ which would be restored in the expressions as and when required. \\

In the case of charged object,
$\Phi(R) = \frac{M - Q^2/2R}{R}$ where electric field energy $Q^2/2R$ lying exterior to radius $R$ is subtracted out from mass $M$. Again $\Phi(R)=1/2$ gives the familiar charged black hole  horizon, $R_+ = M(1 + \sqrt{1 - \alpha^2})\, ,\alpha^2 = Q^2/M^2$.
On the other hand for the Buchdahl star, $\Phi(R)=4/9$ would give
 \cite{dad19, rot07, sum-dad20},
\begin{equation}
M/R = \frac{8/9}{1+\sqrt{1-(8/9)\alpha^2}} \, ,
\end{equation}
which is, $(M/R)_{Buch} = (8/9)(M/R)_{BH} (\alpha^2 \to (8/9)\alpha^2)$.\\

It was rather straightforward to define potential in the static case because of spherical symmetry while it is not so for the axially symmetric rotating case. In this case we have to filter out the inherent frame dragging effect by considering potential for axial motion which entirely lies in the $t-r$ plane. It would then involve only gravitational contribution of rotation. It can be easily seen from the Kerr metric in the standard Boyer-Lindquist coordinates,
\begin{equation}
ds^2= \frac{\Delta}{\rho^2}d\tau^2 - \frac{\rho^2}{\Delta}dR^2 - \rho^2 d\theta^2 - \frac{sin^2\theta}{\rho^2}[(R^2+a^2)d\phi - adt]^2\, ,
\end{equation}
where
\begin{eqnarray}
d\tau &=& dt-asin^2\theta d\phi\, , \Delta=R^2-2MR+a^2 \, , \nonumber\\
\rho^2&=&R^2+a^2cos^2\theta \, .
\end{eqnarray}
Then $\Phi(R)$ for the axial motion would be given by
\begin{equation}
\Phi(R) = \frac{MR}{(R^2+a^2)}=\frac{M/R}{1+\beta^2(M/R)^2}\,\, , \beta^2=a^2/M^2\, .
\end{equation}

Now $\Phi(R) = 1/2$ yields the familiar Kerr black hole horizon,
$R_+ = M(1 + \sqrt{1 - \beta^2})$, while for the rotating Buchdahl star, $\Phi(R)=4/9$ gives
\begin{align}
M/R & = \frac{8/9}{1+\sqrt{1-(8/9)^2\beta^2}} \nonumber \\
& = (8/9)(M/R)_{BH} (\beta^2 \to (8/9)^2\beta^2) \, .
\end{align} \\

Generalizing to Kerr-Newman charged and rotating object described by the above metric with $\Delta = R^2-2MR+a^2+Q^2$, we shall similarly have
\begin{eqnarray}
\Phi(R) &=& \frac{MR - Q^2/2}{(R^2+a^2)}=\frac{M/R - (\alpha^2/2)(M/R)^2}{1+\beta^2(M/R)^2}\, .
\end{eqnarray}
Then $\Phi(r) = 1/2$ gives the familiar K-N black hole horizon, $R_+ = M(1 + \sqrt{1 - \alpha^2 - \beta^2})$ while the corresponding Buchdahl star for $\Phi(R) = 4/9$ we have
\begin{align}
M/R & = \frac{8/9}{1+\sqrt{1- (8/9)\alpha^2 - (8/9)^2\beta^2
}} \nonumber \\
& = (8/9) (M/R)_{BH} (\alpha^2 \to (8/9)\alpha^2, \beta^2 \to (8/9)^2\beta^2) \, .
\end{align}\\

Note that spherically symmetric metric, Schwarzschild or Reissner-Nordstr{\"o}m, describes a general static object which could be a black hole or a non black hole object like a Buchdahl star. In contrast, strictly speaking the Kerr metric describes only a rotating black hole and not a rotating object in general. A rotating object would in general be deformed due to rotation and would in general have multipole moments. That is what the Kerr metric cannot accommodate and hence it can only describe a black hole which has no multipole moments --- no hair. Despite vigorous attempts over decades, there still exists no proper metric describing a non black hole rotating object, one has therefore resort to the Kerr metric for description of rotating Buchdahl star which should be valid only in the first approximation. Despite this the result may however be indicative of a general behavior. \\

We have thus computed the compactness, $M/R$ expression for the various cases which we would now use in the next section to compute the maximum force in these cases.\\

\section{Maximum force}

We compute force between two equal mass objects one of which is charged and rotating while the other is neutral and non-rotating. Their boundaries touch along the axis. The aim is to include only the gravitational contribution of charge and rotation and not electromagnetic and spin-spin interactions. We shall find contributions of charge and rotation to the maximum force. \\

Differentiating $\Phi(R)$ in Eq (11) we obtain % ic   objects of equal mass, charge and rotation parameters by differentiating the expression for  given above in Eq. (8) and so we obtain, %which would be given by %black holes for charged and rotating object employing the Kerr-Newman metric for which the potential is given by,
\begin{equation}
F =
\frac{M^2}{R^2}\frac{1-\alpha^2M/R-\beta^2M^2/R^2}{(1+\beta^2M^2/R^2)^2}\, .
\end{equation}\\

Now the maximum force between two equal mass black holes, one of which is Kerr-Newman and the other Schwarzschild, touching each-other at the horizon on the axis, $\theta=0$, for which $M/R = (1+\sqrt{1-\alpha^2-\beta^2})^{-1}$, would be given by
\begin{align}
F_{max}(KN-BH) & = \frac{\sqrt{1-\alpha^2-\beta^2}}{2(1+\sqrt{1-\alpha^2-\beta^2})-\alpha^2} \nonumber\\
& = 4\frac{\sqrt{1-\alpha^2-\beta^2}}{2(1+\sqrt{1-\alpha^2-\beta^2})-\alpha^2}
\nonumber\\ &\times F_{max}(Sch-BH)\, ,
\end{align}
where $F_{max}(Sch-BH) = c^4/4G$ when $\alpha^2=\beta^2=0$. \\

For the Reissner-Nordstr{\"o}m and Kerr black holes it will respectively read as follows:
\begin{equation}
F_{max}(RN-BH)= 4\frac{\sqrt{1-\alpha^2}}{(1+\sqrt{1-\alpha^2})^2}F_{max}(Sch-BH) \, ,
\end{equation}
and
\begin{equation}
F_{max}(Kerr-BH)= 2\frac{\sqrt{1-\beta^2}}{1+\sqrt{1-\beta^2}}F_{max}(Sch-BH)\, .
\end{equation}\\

Their counterparts for the Buchdahl stars would be obtained by multiplying by $(8/9)^2$ and writing $\alpha^2 \to (8/9)\alpha^2, \beta^2 \to (8/9)^2\beta^2$ in the above expressions; i.e.,
\begin{widetext}
\begin{eqnarray}
F_{max}(KN-Buch) & =& (8/9)^2 F_{max}(KN-BH)(\alpha^2 \to (8/9)\alpha^2, \beta^2 \to (8/9)^2\beta^2) \nonumber \\
& = &4F_{max}(KN-BH)(\alpha^2 \to (8/9)\alpha^2, \beta^2 \to (8/9)^2\beta^2) F_{max}(Buch)\, ,
\end{eqnarray}
\end{widetext}
where $F_{max}(Buch)(\alpha^2=\beta^2=0) = (8/9)^2 F_{max}(Sch-BH)= (8/9)^2 c^4/4G$. \\

In particular maximum force for charged and rotating Buchdahl star would read respectively as follows:
\begin{eqnarray}
F_{max}(RN-Buch) &=& 4\frac{\sqrt{1-(8/9)\alpha^2}}{2(1+\sqrt{1-(8/9)\alpha^2})-\alpha^2}\nonumber\\ &\times & F_{max}(Buch)\, ,
\end{eqnarray}
and
\begin{eqnarray}
F_{max}(Kerr-Buch) &=& 4\frac{\sqrt{1-(8/9)^2\beta^2}}{2(1+\sqrt{1-(8/9)^2\beta^2})}\nonumber\\&\times & F_{max}(Buch)\, ,
\end{eqnarray}\\
where $F_{max}(Buch)= (8/9)^2 F_{max}(Sch-BH)= (8/9)^2 c^4/G$. \\

For $\alpha^2 = 1$, it reduces to $F_{max}(RN-Buch) = (4/5) F_{max}(Buch)$. Similarly for rotating Buchdahl star with $\beta^2=1$, we would have $F_{max}(Kerr-Buch) = 4\sqrt{17}/(9+\sqrt{17})F_{max}(Buch) \approx (8/13)F_{max}(Buch)$. It is interesting to note that extremality for Buchdahl star is over-extremality for black hole. The maximum force vanishes for extremal values which are $\alpha^2=1, \beta^2=1$ for RN and Kerr black holes respectively. But these values are quite fine for Buchdahl star and give non-vanishing value for the maximum force. For Buchdhal star, the corresponding extremal values are $\alpha^2=9/8, \beta^2=(9/8)^2$. \\

Note that $(M/R)_{Buch} = (8/9)(M/R)_{BH} (\alpha^2 \to (8/9)\alpha^2, \beta^2 \to (8/9)^2\beta^2)$, and maximum force is given in terms of $M/R$, that is why maximum force in the two cases is similarly transformed. Maximum force is though independent of mass but it does depend upon charge to mass, $\alpha^2=Q^2/M^2$, and spin to mass, $\beta^2=a^2/M^2$ ratios. \\

\section{Lovelock gravity}

In a $D$-dimensional spacetime, gravity can be described by an action
functional involving arbitrary scalar functions of the metric and curvature,
but not derivatives of curvature. In general, variation of such an arbitrary
Lagrangian would lead to an equation having fourth-order derivatives of the
metric. For them to be of second order, the gravitational lagrangian, $L$,
is constrained to be of the following Lovelock form, \cite{Love}:

\begin{equation}
L=\sum\lim_{N}\alpha_{N}L_{N}^{{}}=\alpha_{N}\frac{1}{2^{N}}\delta
_{c_{1}d_{1}c_{2}d_{2}....c_{n}d_{n}}^{a_{1}b_{1}a_{2}b_{2}...a_{n}b_{n}}R_{a_{1}b_{1}}^{c_{1}d_{1}}R_{a_{2}b_{2}}^{c_{2}d_{2}}....R_{a_{n}b_{n}}^{c_{n}d_{n}},
\label{love1}
\end{equation}%
where $\delta _{rs...}^{pq..}$ is the completely antisymmetric determinant
tensor. Note that $N=1, 2$ respectively correspond to the familiar linear Einstein-Hilbert lagrangian and the quadratic Gauss-Bonnet lagrangian, which is given by

\begin{equation}
L_{2}\equiv L_{GB}=(1/2)(R_{abcd}R^{abcd}-4R_{ab}R^{ab}+R^{2}).
\label{love2}
\end{equation}\\

Lovelock's lagrangian is a sum over $N$, where each term is a homogeneous
polynomial in curvature and has an associated dimensionful coupling
constant, $\alpha_{N}$. Moreover, the complete antisymmetry of the $\delta $
tensor demands $D\geq 2N$, or it would vanish identically. Even for $D=2N$
the lagrangian reduces to a total derivative. Lovelock's lagrangian, $L_{N},$
is therefore non-trivial only in dimension $D\geq 2N+1$. \\

Lovelock theory is the most natural and quintessential higher-dimensional
generalization of GR with the remarkable property that the field equations
continue to remain second order in the metric tensor despite the action
being a homogeneous polynomial in the Riemann tensor. \\

A particular case of interest is that of the \textit{pure Lovelock}
which has only one $N^{th}$-order term in the lagrangian without a sum over lower orders in the action and the equations of motion. It distinguishes itself by the property that gravity is kinematic in all critical odd $D=2N+1$ dimensions. It is well known that GR is kinematic in $D=2\times1+1=3$ in the sense that Riemann is entirely given in terms of Ricci and hence there can exist no no-trivial vacuum solution. Similarly for Lovelock theory, Lovelock Riemann \cite{cam-dad, dgj1} is entirely determined by the corresponding Ricci in all $D=2N+1$, and hence there can occur no vacuum solution. Pure Lovelock theory thus universalizes kinematic property to all critical odd $D=2N+1$ dimensions. \\

Pure Lovelock gravity is kinematic in all critical odd $D=2N+1$ and there can exist no non-trivial vacuum solution unless $D\geq2N+2$.  Finally, variation of the lagrangian with respect to the metric, for pure Lovelock theories, leads to the following second-order equation,

\begin{equation}
-\frac{1}{2^{N+1}}\delta
_{c_{1}d_{1}c_{2}d_{2}....c_{n}d_{n}}^{a_{1}b_{1}a_{2}b_{2}...a_{n}b_{n}}R_{a_{1}b_{1}}^{c_{1}d_{1}}R_{a_{2}b_{2}}^{c_{2}d_{2}}....R_{a_{n}b_{n}}^{c_{n}d_{n}}=8%
\pi GT_{ab}.  \label{love3}
\end{equation}

Since no derivatives of curvature appear, this equation is of second order
in derivatives of the metric tensor. Although not directly evident, the
second derivatives also appear linearly and the equations are therefore
quasi-linear, thereby ensuring unique evolution. \\

Another property that singles out pure Lovelock is the existence of \emph{bound orbits} around a static object \cite{dgj2}. Note, that in GR,
bound orbits exist around a static object only in $D=4$. In view of these remarkable features, it has been argued that pure Lovelock is an attractive gravitational equation in higher dimensions \cite{dad16}. \\

As with the Schwarzschild solution for GR, there exists an exact solution for a pure Lovelock black hole \cite{dpp}, and it is given by Eq.~(5), with
\begin{equation}
\Phi(R)=\frac{GM}{R^n}, \quad n =\frac{(D-2N-1)}{N}
\label{z}
\end{equation}
where $G$ is the gravitational constant appropriate for the corresponding dimension and Lovelock degree $N$. Clearly $D\geq2N+2$ for non-trivial vacuum solution, and black hole horizon is given by
\begin{equation}
\Phi(R)=\frac{GM}{R^n} = \frac{1}{2} \, ,
\end{equation}
yielding horizon, $R_H^n = 2GM/c^2$. \\

The force between two neutral static pure Lovelock black holes with their horizons touching would be given by
\begin{equation}
F= n \frac{GM^{2}}{R_H^{n +1}}.
\end{equation}
This would not be independent of black hole masses unless $n=(D-2N-1)/N =1; i.e., D=3N+1$. Then the maximum force takes the same value as for GR in $D=4$; i.e,
\begin{equation}
F_{max}(BH) = \frac{c^4}{4G}\, .
\end{equation}

With $n=1$ equivalently $D=3N+1$, Buchdahl star is identified by $\Phi(R)=4/9$ and the maximum force between two equal mass Buchdahl stars is
\begin{equation}
F_{max}(Buch) = (8/9) F_{max}(BH) \, ,
\end{equation}
where $F_{max}(BH)= c^4/4G$.\\

The pure Lovelock analogue of force in the Planck unit \cite{JB-ND} would read as
\begin{equation}
F_{pl}=G^{2/(2-D)}c^{(4+D)/(D-2)}h^{(3N+1-D)/(2-D)}\, ,
\end{equation}
which would be free of $h$ if and only if $D=3N+1$. \\

The property that the maximum force is universal and is equal to  $c^4/4G$ uniquely picks out pure Lovelock theory in $D=3N+1$ which includes GR for $N=1$ and $D=4$. \\

\section{Discussion}

Existence of maximum force depends upon occurrence of black hole horizon that marks the lower bound on distance separation between the two black holes. Further its independence on mass of black holes critically hinges on the inverse square law which singles out four dimension in GR. For the inverse square law, potential should go as $1/R$ which in GR could only happen for $D=4$ because potential  goes as $1/R^{D-3}$. In contrast in pure Lovelock gravity it goes as
$1/R^n$ \cite{dpp} where $n=(D-2N-1)/N$ which would be unity for $D=3N+1$ \cite{sum-dad}. That is, force would be inverse square for $D=4, 7, 10, ...$ respectively for linear GR, $N=1$, quadratic Gauss-Bonnet, $N=2$, cubic, $N=3$, and so on. Force would therefore be inverse square for the entire dimensional spectrum, $D=3N+1$. \\

We could paraphrase this general result on the lines of Bertrand's theorem of classical mechanics as follows:

{\it The property that maximum force for uncharged static black hole and Buchdahl star is entirely given in terms of the fundamental constant velocity of light and the gravitational constant, uniquely picks out pure Lovelock gravity in the dimensional spectrum $D=3N+1$.} \\

It is universal and also free of the Planck's constant in pure Lovelock gravity in $D=3N+1$ as we have seen in Eqs (3, 28). This indicates that this is purely classical result which would remain true even when quantum gravity effects are included or in full theory of quantum gravity. It indicates something fundamental and natural about these non-quantum universal units which is unique to $D=4$ in GR and in general to $D=3N+1$ in pure Lovelock gravity. \\

It is well known that GR is kinematic in three dimension because Riemann is entirely given in terms of Ricci, that there can exist no non-trivial vacuum solution. Pure Lovelock gravity universalizes this property that it is kinematic (Lovelock Riemann is given in terms of Lovelock Ricci and there could occur no non-trivial pure Lovelock vacuum solution) in all critical odd $D=2N+1$ dimensions \cite{cam-dad, dgj1}. The action Lagrangian contains only one $N$th order term without sum over lower orders. There are several interesting features of pure Lovelock gravity, for example bound orbits, which in GR exist only for $D=4$, could exist in the dimensional window $3N+1 \leq D \leq 4N$ \cite{dgj2, dad16}. Existence of maximum force given entirely in terms of $c$ and $G$ adds yet another distinguishing property for pure Lovelock gravity. We have elsewhere argued \cite{dad16} that pure Lovelock equation is perhaps the right gravitational equation in higher dimensions.  \\

We have evaluated the maximum force for black holes and Buchdahl stars having charge and/or rotation. It is interesting to note how the compactness ratio, $M/R$ and maximum force transform in going from one to the other. That is, the compactness ratio, $(M/R)_{Buch} = (8/9)(M/R)_{BH}$, and maximum force, $F_{max}(Buch) = (8/9)^2 F_{max}(BH)$ with $\alpha^2 \to (8/9)\alpha^2$ and $\beta^2 \to (8/9)^2\beta^2$. In particular for the neutral static Buchdahl star, $(M/R)_{Buch} = (8/9)(M/R)_{BH}= 8/9 \times 1/2 = 4/9$ and $F_{max}(Buch)= (8/9)^2 F_{max}(BH)$ where $F_{max}(BH) = c^4/4G$. For uncharged static Buchdahl star, compactness ratio is scaled relative to black hole by the factor $8/9$ while the maximum force by its square, and $\alpha^2$ and $\beta^2$ are similarly scaled respectively. \\

Also note that the same factor $8/9$ is the ratio between square of  their escape velocities; i.e., $(v_{Buch}/v_{BH})^2 = 8/9$ and this is universal \cite{sum-dad22} as it is the same for charged and/or  rotating objects as well. Further it turns out that like the black hole \cite{nar-dad}, non-extremal charged Buchdahl star cannot also  be extremalized. It is because the parameter window for particles reaching the star pinches off as extremality is approached. This is exactly parallel to what happens for black hole. Thus black hole and Buchdahl star share good bit of similar features. \\

Maximum force or maximum luminosity could be computed for non black hole object, Buchdahl star because it has like black hole a bound on $M/R$. Observationally luminosity of any phenomenon involving compact stars would always be bounded by $(8/9)^2$ of black hole luminosity. The most remarkable feature is that the bound is given in terms of the fundamental constant velocity of light and the gravitational constant.

\section{Acknowledgements} This work is supported by the Chinese Academy of Sciences President's International Fellowship Initiative Grant No. 2020VMA0014. I warmly thank Sanjar Shaymatov for help in preparing the manuscript. \\

%\bibliographystyle{apsrev4-1}
%\bibliography{gravreferences}

\begin{thebibliography}{99}

\bibitem{Gibb2002} G.W. Gibbons, Found.\ Phys.\ \textbf{32},189 (2002)

\bibitem{JBGib} J.D. Barrow and G.W. Gibbons, Mon.\ Not.\ Roy.\ Astron.\
Soc.\ \textbf{446}, 3874 \ (2014)

\bibitem{dyson} F. Dyson, in \textit{Interstellar Communication}, ed.
Cameron A.G., Benjamin Inc., New York, (1963) chap 12

\bibitem{schiller} C. Schiller, Int.\ J.\ Theor.\ Phys.\ \textbf{45}, 221 (2006)

\bibitem{sperhake} U. Sperhake, E. Berti~and V. Cardoso, Comptes Rend. Phys. \textbf{14}, 306 (2013)

\bibitem{moffatt} J. W. Moffat, JCAP \textbf{0603}, 004 (2006)

\bibitem{JB-ND} J. D. Barrow, N. Dadhich, Phys. Rev. {\bf D102}, 064018 (2020): arxiv:2006.07338.

\bibitem{Love} D. Lovelock, J. Math. Phys.\textbf{12}, 498 (1971)

\bibitem{schiller21} C. Schiller, arxiv:2112.15418.

\bibitem{ong} S. di Gennaro, M. R. R. Good, Y. C. Ong, arxiv:210813435.

\bibitem{ata} K. Atazadeh, Phys. Lett. {\bf B820}, 136590 (2021); arxiv:2111.00212.
    
\bibitem{mirza} B. Mirza, Z. Mirzaiyan, H. Nadi, Ann. Phys. {\bf 415}, 168117 (2020). 

\bibitem{Barrow2004} J.D. Barrow, Class. Quantum Gravity \textbf{21}, L79 (2004)

\bibitem{Barrow2004a} J.D. Barrow, Class. Quantum Grav. \textbf{21}, 5619 (2004)

\bibitem{Barrow2004b} J.D. Barrow and C.G. Tsagas. Class. Quantum Grav. \textbf{22}, 1563 (2005)

\bibitem{BarGib2} J.D. Barrow and G.W. Gibbons, Phys. Rev. \textbf{D95}, 064040 (2017)

\bibitem{JBsing} J.D. Barrow, Int. J. Mod. Phys. {\bf D29}, 14, 2043008 (2020); arXiv: 2005.06809.

\bibitem{JBnewt} J. D. Barrow, Class. Quantum Gravity \textbf{37}, 125007 (2020)

\bibitem{sum-dad} S. Chakraborty and N. Dadhich, Euro. Phys. J. \textbf{C78}, 81 (2018).

\bibitem{buch} H. A. Buchdahl, Phys. Rev. {\bf 116}, 1027 (1959).

\bibitem{andrea08} H Andreasson, J. Diff. Equations {\bf 245}, 2243 (2008).

\bibitem{kara-stal08} P Karageorgis, J Stalker, Class. Quant. Grav. {\bf 25} (2008) 195021.

\bibitem{mak-stuch} M K mak, P N Dobson, Jr, T Harko, Mod, Phys. Lett. {\bf A15}, 2153 (2000);
T Harko, M K Mak, J. Math. Phys. {\bf 41}, 4752 (2000); H Andreasson, C G Boehmer, A Mussa, Class. Quant. Grav.
{\bf 29}, 095012 (2012), arxiv:1201.5725; Z Stuchlik, Acta. Phys. Slov. {\bf 50}, 219 (2000), arxiv:0803.2530.

\bibitem{andrea09} H Andreasson, Commun. Math. Phys. {\bf 288}, 715 (2009); gr-qc/0804.1882.

\bibitem{alho} A. Alho, J. Natario, P. Pani, G. Raposo, arxiv:2202.00043.

\bibitem{dad19} N. Dadhich, JCAP{\bf04}, 035 (2020); arxiv:2019.03436.

\bibitem{sum-dad22} S. Chakraborty, N. Dadhich, Universality of escape velocity, to be submitted.

\bibitem{rot07} A. Giuliani, T. Rothman, Gen. Relativ. Grav. {\bf39}, 757 (2007); gr-qc/0702078.

\bibitem{sum-dad20} S. Chakraborty, N. Dadhich, Phys. Dark Univ. {\bf30}, 100658 (2020); arxiv:2005.07504.

%\bibitem{sporea} C.A. Sporea, Mod. Phys. Lett. A in press 1812.09945

%\bibitem{Barclif} J.D. Barrow and T. Clifton, Phys. Rev. \textbf{D72},103005 (2005)

%\bibitem{clifton } T. Clifton, \textit{Alternative Theories of Gravity}, Ph.D. thesis, Univ. Cambridge, (2006) https://arxiv.org/abs/gr-qc/0610071

%\bibitem{BD} C.H. Brans and R.H. Dicke, Phys. Rev. \textbf{124}, 925 (1961)

%\bibitem{camp} M. Campanelli and C.O. Lousto, Int. J. Mod. Phys. \textbf{D2}, 451 (1993)

%\bibitem{bad} A. Bhadra and K. Sarkar, Gen. Rel. Grav. \textbf{37}, 2189
(2005)

%\bibitem{swh} S.W. Hawking, Commun. Math. Phys. \textbf{43}, 199 (1975)


\bibitem{cam-dad} X. Camanho and N. Dadhich, Euro. Phys. J. \textbf{C76},149 (2016) {}

\bibitem{dgj1} N. Dadhich, S. Ghosh and S. Jhingan, Phys. Lett. {\bf B711}, 196 (2012)

\bibitem{dgj2} N. Dadhich, S. Ghosh and S. Jhingan, Phys. Rev. {\bf D88}, 124040 (2013)

\bibitem{dad16} N. Dadhich, Euro. Phys. J. \textbf{C76}, 104 (2016)

\bibitem{dpp} N. Dadhich, J. Pons and K. Prabhu, Gen. Relativ. Grav. \textbf{45}, 1131 (2013)

\bibitem{nar-dad} K. Narayan, N. Dadhich, Phys. Lett. {\bf A231}, 335 (1997); grqc/9704070.




\end{thebibliography}

 \end{document}